\documentclass[linenumbers, twocolumn]{aastex631}
 \usepackage{ifpdf}
 \usepackage{graphicx, subfigure}
 \usepackage{natbib}
 \usepackage{times}
 \usepackage{amsmath}
 \usepackage{textcomp}
 \usepackage{lineno}
 \usepackage[colorinlistoftodos,prependcaption,textsize=tiny]{todonotes}
 \usepackage{xcolor}
 \usepackage{gensymb}
\usepackage{comment}

\newcommand{\bbe}{\begin{eqnarray}}
\newcommand{\ee}{\end{eqnarray}}
\newcommand{\n}{\nonumber}

\begin{document}
\nolinenumbers

\title{The Gamma-ray Luminosity Function of Flat-Spectrum Radio Quasars}

\author{Garima Rajguru}
\affil{Department of Physics and Astronomy, Clemson University, Kinard Lab of Physics, Clemson, SC 29634-0978, USA}
\email{grajgur@clemson.edu}

\author{L. Marcotulli}
\altaffiliation{NHFP Einstein Fellow}
\affil{Department of Physics, Yale University, PO Box 208120, New Haven, CT 06520-8120, USA}
\affil{Yale Center for Astronomy and Astrophysics, New Haven, CT 06520, USA}

\author{M. Ajello}
\affil{Department of Physics and Astronomy, Clemson University, Kinard Lab of Physics, Clemson, SC 29634-0978, USA}

\author{M. Di Mauro}
\affil{Istituto Nazionale di Fisica Nucleare, Sezione di Torino, Via P. Giuria 1, 10125 Torino, Italy}

\author{M. Urry}
\affil{Department of Physics, Yale University, PO Box 208120, New Haven, CT 06520-8120, USA}
\affil{Yale Center for Astronomy and Astrophysics, New Haven, CT 06520, USA}


\begin{abstract}
\nolinenumbers

We have utilized the largest sample of $\gamma$-ray selected \textit{Fermi} flat-spectrum radio quasars (FSRQs) ever used (519 sources) to construct the luminosity function and its evolution through the cosmic history. In addition to spanning large redshift ($0<z\lesssim 4$) and luminosity ranges ($2.9\times10^{43}$\,erg s$^{-1}$ -- $7.3\times10^{48}$\,erg s$^{-1}$), this sample also has a robust calculation of the detection efficiency associated with its observation, making its selection effects and biases well understood. We confirm that the local luminosity function is best explained by a double power law. The evolution of the luminosity function of FSRQs follows a luminosity-dependent density evolution. FSRQs experience positive evolution with their space density growing with increasing redshift up to a maximum redshift, after which the numbers decrease. This peak in redshift occurs at larger redshifts for higher luminosity sources and at lower redshifts for lower luminosity sources. We find an unexpected similarity between the luminosity function of FSRQs and that of BL Lacertae objects at intermediate luminosity. This could be a sign of a strong genetic link between the two blazar sub-classes or that BL Lac samples are contaminated by large amounts of FSRQs with their jets nearly perfectly aligned with our line of sight.

\end{abstract}



\section{Introduction} \label{sec:intro}
Blazars are a subclass of active galactic nuclei (AGNs), which have their relativistic jet oriented along our line of sight \citep{1995Urry}. Blazar jets are relativistically beamed, since they are pointed at us, and their spectral energy distribution spans a broad energy range from radio waves to gamma-rays \citep[see e.g.,][]{Padovani2017_Review}. The energetic electrons present in the jet give rise to synchrotron radiation, with its emission spanning from radio to UV or X-ray energies. The high-energy emission (generally in the keV-GeV range) is considered to be due to the inverse Compton scattering of photons coming from the jet (or coming from external sources) by the energetic jet electrons, upscattering the former to much higher energies \citep[see e.g.,][]{Blandford2019_JetReview}. Blazars that are devoid of strong emission lines (equivalent width; $EW<5$\AA) in their optical spectra are called BL Lacertae objects (BL Lacs), whereas objects showing the broad emission lines are called flat-spectrum radio quasars  \citep[FSRQs; ][]{1995Urry}; together they form the two subcategories of blazars.

The evolution of blazars encodes the change in a particular property of blazars (such as luminosity or number distribution) with change in redshift. The quest for the luminosity function (LF) of quasars started with optical and radio samples \citep{Schmidt1968}, comprising mainly of flat-spectrum quasars owing to their large bolometric luminosities. Several studies have investigated the evolution of blazars in different energy bands, such as in radio \citep{Peacock1985,DunlopPeacock1990,Wall2005}, soft X-ray \citep{Padovani1992,GiommiPadovani1994,Rector2000,WolterCelotti2001,Caccianiga2002,Beckmann2003,Padovani2007}, hard X-ray \citep{Giommi1991,Ajello2009,Toda2020,Lea2022} and gamma-rays \citep{Chiang1995,ChiangMukherjee1998,Ajello2012,Ajello2014}. FSRQs seem to show a positive evolution (they were more numerous in the past) until a redshift cut-off which depends on luminosity \citep{DunlopPeacock1990,Padovani2007,Wall2008,Ajello2009,Ajello2012}. BL Lacs, on the other hand, show mixed evolutionary trends including positive, negative \citep{Rector2000,Beckmann2003} and no evolution at all \citep{Caccianiga2002,Padovani2007}. Limited sample size, biases in data and redshift incompleteness have been major challenges in the determination of the LF of BL Lacs.

In the $\gamma$-ray regime, the attempt to constrain the LF of blazars started with the EGRET sample \citep{Chiang1995, ChiangMukherjee1998,MuckePohl2000,Dermer2007,Bhattacharya2009}; however, they suffered from small sample sizes ($\sim 60$) and redshift incompleteness. Prior to this, the $\gamma$-ray LF was derived from the LF of X-ray or radio bands, following their correlations with $\gamma$-ray luminosities \citep[e.g.,][]{SteckerSalamon1996,NarumotoTotani2006,InoueTotani2009,SteckerVenters2011}, leading to large uncertainties in the models.

There are major open-ended questions that the evolution of blazars can shed light on:
\begin{enumerate}
    \item Evolution of supermassive black holes - At the heart of large galaxies resides a supermassive black hole \citep[SMBH; $M_{BH} > 10^8 M_\odot$ ;][]{Ghez1998}, which is considered to be the underlying engine of AGNs \citep{Lynden-Bell1969}. Blazars, as would be expected, are found to reside in large elliptical galaxies \citep[e.g.,][]{Urry1999,Falomo2000,Scarpa2000,ChiabergeMarconi2011,Olguin2016}. They can be observed up to large redshifts \citep[$z>2$; ][]{Romani2006,Ajello2009,Marcotulli2020}, owing to their high luminosities. Thus, the evolution of blazars can be used as tracers for the formation and evolution of very heavy SMBHs \citep[e.g.,][]{Volonteri2011,Sbarrato2015,BlazarEpochReionize2024}. Indeed, the growth of SMBH in the early universe remains an open question.
    
    Since blazars are found in old massive ellipticals, which are thought to have undergone mergers in their lifetime, it is anticipated that merger events may be a potential avenue of fueling strong accretions and powering jets \citep{BertiVolonteri2008,Mayer2010,Volonteri2010,Chiaberge2015,Paliya2020}. Thus, mapping the evolution of jetted AGNs, such as blazars, give us an idea about merger activity through cosmic time.
    \item Contribution of blazars to the extragalactic gamma-ray background (EGB) - The universe has an all-sky $\gamma$-ray glow known as the EGB, which was first observed by the Small Astronomy Satellite \citep[SAS-2; ][]{Fichtel1975}. The EGB comprises of three main components arising from the resolved sources, the unresolved sources and the truly diffused processes. The EGB from 100 MeV to 820 GeV has been measured using \textit{Fermi} data \citep{Ackermann2015_IGRB}. Since blazars are the most numerous source class detected by \textit{Fermi}, it is not surprising that the unresolved blazars will have a large contribution to the EGB as well \citep{2015Ajello,Lea2020}. Knowing the LF of blazars more accurately enables us to better account for those undetected sources and determine their contribution to the EGB.
\end{enumerate}

\citet{Ajello2012} derived the LF of \textit{Fermi} FSRQs using a sample of 186 objects, in the energy range of 100 MeV to 100 GeV, originating from the first year of operation of the \textit{Fermi} Large Area Telescope (LAT).

In this work, we use the largest dataset of FSRQs (519 sources) ever used to study the  LF at $\gamma$-rays  from 100 MeV to 1 TeV. This is possible due to the uniform sky survey by \textit{Fermi}-LAT and multiple campaigns to determine the redshift of these sources. 
This paper is organized in the following manner. In Section \ref{sec:sample}, we discuss the target selection and dataset. In Section \ref{sec:analysis}, we elaborate on the analysis technique employed to determine the LF of our source sample. Sections \ref{sec:results} and \ref{sec:conc} provide the results of our analysis and the corresponding discussions and implications. Throughout this paper, we use a flat $\Lambda$CDM cosmological model with $H_0=67 \rm ~km~s^{-1}~Mpc^{-1}$, $\Omega_m = 0.3$ and $\Omega_{\Lambda}=0.7$.

\section{The Sample} 
\label{sec:sample}

\begin{table}
\begin{center}
\caption{Composition of the Gamma-Ray Blazar Sample \\ ($|b|> 20^\circ$).}
\begin{tabular}{lcr}
\hline
\hline
Class  &  No. of Objects &Available Redshifts$^a$\\ 
\hline
\hline
Total & 2680 & 1365 \\
\tableline
\tableline
BL Lacs & 1035 & 705 \\
FSRQs & 519 & 519\\
BCUs & 561 & 102\\
Others$^{b}$ & 41 & 39\\
Unassociated& 524 & 0\\
\tableline
\tableline
Redshift Estimate Available  &  1365 & \\
Redshift Estimate Unavailable  &  1315 & \\
\hline
\hline
\multicolumn{3}{l}{
\begin{minipage}{\linewidth} 
\tablenotetext{a}{or spectra with identified features.}
\tablenotetext{b}{Other objects include radio galaxies, narrow-line Seyfert 1 objects, AGNs, etc.}
\end{minipage}
}
\end{tabular}
\label{Tab:sample}
\end{center}
\end{table}

The sample used is the one derived in \citet{Lea2020} that used eight years of Pass 8 events \citep[P8R3;][]{Bruel2018} of the $Fermi$ Large Area Telescope \citep[LAT; ][]{LAT_Atwood2009}. \citet{Lea2020} used it to derive the intrinsic source-count distribution of the sample. The catalog consists of 2680 sources detected above the Galactic latitude of $|b|> 20^\circ$ and between the energy range of 100 MeV to 1 TeV. \citet{Lea2020} derived the intrinsic source-count distribution associated with this catalog down to $\sim 10^{-10} \rm \, ph\, cm^{-2}\, s^{-1}$ ($\sim 10^{-12} \rm \,  erg\, cm^{-2}\, s^{-1}$), which is an order of magnitude lower than any previous measurement. The efficiency measurement reported in \citet{Lea2020}, computed through Monte Carlo simulations, ensures that the survey and detection biases are taken into account. Moreover, the analysis of \citet{Lea2020} was performed while taking into account the possible curvature of the spectral energy distribution (SED) of blazars, making it the most robust calculation of efficiency to date for a \textit{Fermi}-LAT sample.


Since for our analysis we need a sample of blazars with known redshift, we cross-match \citet{Lea2020}'s sample with the fourth LAT AGN catalog -  Data Release 3 \citep[4LAC-DR3; ][]{4lac_dr3} employing the Bayesian cross-matching tool NWAY\footnote{https://github.com/JohannesBuchner/nway} \citep{Salvato_2018_nway}. Based on the positions and positional uncertainties of the sources in these two catalogs, taking into account the different covered sky areas, NWAY returns both the probability of any possible counterpart for a given source ($p_{any}$) and the relative posterior probabilities of the possible associations ($p_i$). We only keep sources with an associated counterpart that satisfies the requirements of $p_{any}>95\%$ and $p_i>95\%$. This results in 2156 secure 4LAC-DR3 associations with the sample of \citet{Lea2020}.

Of these sources, 519 are FSRQs, which we use in this work. 
This represents the largest sample to date that has been used to compute the LF of FSRQs. The source classifications and redshifts of these objects were updated using the latest data release of 4LAC \citep[4LAC-DR3; ][]{4lac_dr3} and the information available on the website ZBLLAC - A Spectroscopic Library of BL Lac Objects\footnote{\href{https://web.oapd.inaf.it/zbllac/index.html}{https://web.oapd.inaf.it/zbllac/index.html}}. Moreover, redshifts were also updated from the recent optical spectroscopy studies \citet{Goldoni2021} and \citet{Redshift_OptSpec_SouthHem}. Redshifts determined using photometric observations were also included from studies such as \citet{kaur2017,Yong2024,Yong2024arXiv}. All 519 FSRQs have measured redshifts and the highest redshift in our sample is $z=3.65$.

Table \ref{Tab:sample} reports the composition of our sample in terms of source types and redshift information. As we can see, 1035 of the 2680 sources are BL Lacs, 519 are FSRQs and 561 sources are blazar candidates of uncertain type (BCUs). 
The incompleteness of a sample is defined as the fraction of objects that are unclassified with respect to the total sample. The incompleteness of our sample is 19\% (0.19). In order to account for the potential FSRQs among the unclassified sources in the sample, we multiply the efficiency by the completeness (0.81) of the sample. This assumes that 24\% of the unclassified sources are FSRQs (a similar ratio of FSRQs to the total number of classified sources).

\section{Analysis} \label{sec:analysis}
\subsection{Form of the Luminosity Function} \label{LF}
The LF of a source class is defined as the number of sources per unit comoving volume ($dV$) and luminosity interval ($dL_\gamma$). We extend this definition to include the $\gamma$-ray photon index ($\Gamma$) of the sources. Therefore, we arrive at the number density of FSRQs as a function of $L_\gamma$, $z$ and $\Gamma$, defined as:
\bbe
\label{eq:1}
\Phi(L_{\gamma},V(z)) = \frac{d^2N}{dL_{\gamma} dV}
\ee
\bbe
\label{eq:2}
\frac{d^3 N}{dL_{\gamma} dz d\Gamma} = \Phi(L_{\gamma},V(z)) \times \frac{dN}{d\Gamma} \times \frac{dV}{dz}
\ee
In Eqn. \ref{eq:2}, $\Phi(L_{\gamma},V(z))$ is the LF, $L_\gamma$ is the rest-frame luminosity between 100 MeV and 1 TeV, and $\frac{dV}{dz}$ is the comoving volume element per unit redshift and per unit solid angle. We have used the prescription given in \citet{Hogg1999} to compute $\frac{dV}{dz}$ ($=\frac{dV_C}{dzd\Omega}$). 

The intrinsic photon index distribution $\frac{dN}{d\Gamma}$ is assumed to be a Gaussian function:
\bbe
\label{eq:3}
\frac{dN}{d\Gamma} = e^{-\frac{ (\Gamma-\mu(L_\gamma))^2}{2\sigma^2}}
\ee
where $\mu(L_\gamma)$ is the luminosity-dependent mean of the Gaussian and $\sigma$ is the dispersion. The luminosity dependence of the photon index has previously been suggested by a number of studies \citep{Ghisellini2009_BlazarDiv, Meyer2012_PLIndexLum, Ajello2014}.

In this work, we follow the parametrization given in \citet{Ajello2014}:
\bbe
\label{eq:4}
\mu(L_\gamma) = \mu^* + \beta \times (\log_{10} (L_\gamma ) - 46)
\ee

We aim to fit a parametric function of the LF, including its evolution, simultaneously. Previous studies inform us about the possible functional form of the LF and its density evolution. \citet{Schmidt1968} and \citet{Lynden-Bell_C_1971} found that the radio and optical LF increased toward fainter luminosities, respectively. \citet{Marshall1983} had parametrized the local LF ($z$ = 0) as a power law, with the evolution following a pure luminosity (only the luminosity changes with time) or a pure density evolution (only the number changes with time). Studies of the LF of beam-dominated sources found that there is a flattening of the slope of the LF at lower luminosities, leading to a double power-law shape \citep[see e.g.][]{UrryShafer1984,Urry1991,Ajello2012}. This was attributed to the beaming of the relativistic jet. In $\gamma$-rays, \citet{ChiangMukherjee1998} found that the local LF for a blazars followed a double power-law shape. Moreover, in hard X-rays, \citet{Ueda2003} and \citet{Lea2022} found that the double power-law shape improved the local LF for AGNs and blazars over a simple power-law.


Considering the above, in our work, the local LF at $z=0$ is described by a smoothly joined double power law \citep[see e.g.,][]{Ueda2003, Hasinger2005, Ajello2012}, combined with the photon index distribution.
\bbe
\label{eq:5}
\Phi(L_{\gamma},V(z=0),\Gamma) =  \n\\
\frac{A}{\ln(10)L_{\gamma}} \left[\left(\frac{L_{\gamma}}{L_{*}}\right)^{\gamma_1}+ \left(\frac{L_{\gamma}}{L_{*}}\right)^{\gamma2} \right]^{-1}
&\times \frac{dN}{d\Gamma}
\ee
where $A$ is the normalization factor, $L_{*}$ is the knee of the LF, $\gamma_1$ and $\gamma_2$ are the power-law indices of the low and high-end luminosities.

\subsection{Parametrization of Evolution} \label{evolution}
The pure luminosity and density evolution in \citet{Marshall1983} was tested using 2 different forms in redshift ($z$): a power-law form ($(1+z)^k$) and an exponential form ($e^{k\tau(z)}$), with $\tau (z)$ being the fractional look-back time. Using a sample of 448 X-ray selected AGNs, \citet{DellaCeca1992} found the best-fit evolution to be a luminosity-dependent density evolution, as suggested earlier by \citet{SchmidtGreen1983,SchmidtGreen1986}. The luminosity-dependent density evolution model proved to be a better fit for both soft and hard X-ray studies of AGNs \citep[e.g. ][]{Miyaji2001,Ueda2003,Hasinger2005}. \citet{Wall2005} showed that, for submillimeter galaxies, the increase and subsequent decrease of space density with increasing redshift is better explained by a power-law combined with an exponential ``roll-off" term. In $\gamma$-rays, \citet{NarumotoTotani2006, Ajello2012, Ajello2014} studied the luminosity-dependent density evolution in blazars.

In order to investigate the evolution of blazars, we use a compound function consisting of the local LF (at $z=0$) and an evolutionary factor ($e(z,L_\gamma)$) dependent on redshift \citep[see e.g.,][]{Wall2008, Ajello2009, Ajello2012, Ajello2014, Lea2022}. We test three different functions to constrain the evolution:
\begin{itemize}
    \item Primarily Density Evolution (PDE), where the number of sources increases or decreases with redshift. It is quantitatively described by:
    \bbe
    \label{eq:6}
    \Phi(L_{\gamma},V(z),\Gamma) &= \Phi(L_{\gamma},V(z=0),\Gamma) \n\\
&\times e(z,L_\gamma)
    \ee
    \item Primarily Luminosity Evolution (PLE), where the luminosity of sources changes with redshift. It is quantitatively described by:
    \bbe
    \label{eq:7}
    \Phi(L_{\gamma},V(z),\Gamma) = \Phi(L_{\gamma}/e(z,L_\gamma),V(z=0),\Gamma)
    \ee
    \item Luminosity-Dependent Density Evolution (LDDE), where the variation (increase/decrease) in the number of sources depends additionally on luminosity, as we look into larger redshift. It is quantitatively described by:
    \bbe
    \label{eq:8}
    \Phi(L_{\gamma},V(z),\Gamma) &= \Phi(L_{\gamma},V(z=0),\Gamma) \n\\
&\times e'(z,L_\gamma)
    \ee
\end{itemize}

The evolutionary factors are given by:
\bbe
\label{eq:9}
e(z,L_\gamma) = (1+z)^k e^{z/\xi}\\
k = k^* + \tau \times (\log_{10} (L_\gamma ) - 46) \n
\ee
where $z$ is the redshift, $k$ is the luminosity-dependent redshift index, $\tau$ is the parameter that relates the redshift index with the luminosity, and $\xi$ the evolutionary cutoff term; and
\bbe
\label{eq:10}
e'(z,L_\gamma) = \left[\left( \frac{1+z}{1+z_c(L_{\gamma})}\right)^{-p1} + \left( \frac{1+z}{1+z_c(L_{\gamma})}\right)^{-p2} \right]^{-1} 
\ee
\bbe
\label{eq:alpha}
z_c(L_{\gamma}) = z_c^*\cdot (L_{\gamma}/10^{48})^{\alpha} 
\ee
Here, $z_c(L_{\gamma})$ corresponds to the (luminosity-dependent) redshift where the evolution gets inverted (positive to negative), with $z_c^*$ being the redshift peak for a FSRQ with a luminosity of $10^{48}$ erg/s. The parameter $\alpha$ is the power law index relating the redshift peak of evolution and the luminosity. The LDDE model has a total of 10
free parameters whereas PLE and PDE have 9 parameters each. We remark that all three models (including PDE and PLE) have a luminosity-dependent density evolution built in the evolutionary factor $e(z,L_\gamma)$.

\subsection{Maximum Likelihood Fit}
The $1/V_{\rm max}$ method in \citet{Schmidt1968} has been commonly employed to measure LFs. However, the redshift binning in this method can bias the results if there is considerable evolution within the bins, especially when the sample spans a large range in luminosity and redshift. Given our large dataset spanning large ranges in $L_\gamma$ and $z$, we employ the maximum likelihood (ML) approach, introduced by \citet{Marshall1983} and used in \cite{Ajello2009,Ajello2012}, to determine the best-fit LF. 

In this method, the luminosity, redshift and photon index space is finely parsed into intervals of size $dL_{\gamma}dzd\Gamma$, such that only 0 or 1 object can occupy each interval. The best-fit LF is obtained by utilizing an ML estimator to compare the number of observed objects found in a certain interval to the number of expected objects (for a given LF model). The expected number of sources in each interval is given by:
\bbe
\label{eq:11}
\lambda(L_{\gamma},z, \Gamma)dL_{\gamma}dz d\Gamma   = \Phi(L_{\gamma},z) \Omega(L_{\gamma},z,\Gamma) \n\\
\times \frac{dN}{d\Gamma}  \frac{dV}{dz} dL_{\gamma} dz d\Gamma
\ee
where $\Omega(L_{\gamma},z,\Gamma)$ is the sky coverage, which represents the probability of 
detecting in this survey a blazar with luminosity $L_{\gamma}$, redshift $z$ and photon index $\Gamma$ (for details see Sec. \ref{sec:SkyCoverage}). This factor accounts for survey biases and selection effects, and was derived for the sample used here by \cite{Lea2020}.

The likelihood function ($L$) is based on joint Poisson probabilities for the detection of 1 source in the $i^{th}$ interval and 0 sources in all other intervals:
\begin{align}
\label{eq:12}
L = & \prod_i \lambda(L_{\gamma,i},z_i,\Gamma_i) dL_{\gamma} dz  d\Gamma e^{-\lambda(L_{\gamma,i},z_i,\Gamma_i) dL_{\gamma} dz d\Gamma} \n\\
& \times \prod_j e^{-\lambda(L_{\gamma,j},z_j,\Gamma_j) dL_{\Gamma} dz d\Gamma}
\end{align}
For ease of computation, we transform the above to the standard expression $S=-2\ln\ L$ and drop the model-independent terms:
\begin{align}
S = & -2 \sum_i \ln \left( \frac{d^3 N}{dL_{\gamma} \, dz \, d\Gamma} \right) \n\\
    & + 2 \int_{\Gamma_{\min}}^{\Gamma_{\max}} \int_{L_{\gamma,\min}}^{L_{\gamma,\max}} 
    \int_{z_{\min}}^{z_{\max}} \lambda(L_{\gamma}, \Gamma, z) \, dz \, dL_{\gamma} \, d\Gamma
\label{eq:13}
\end{align}
Unless specified otherwise, the integration limits are $L_{\gamma,min}$ = $2.9\times10^{43}$\,erg s$^{-1}$, $L_{\gamma,max}$ = $7.3\times10^{48}$\,erg s$^{-1}$, $z_{min}$ = 0.0001, $z_{max}$ = 5.0, $\Gamma_{min}\,=$ 1.0 and $\Gamma_{max}\,=$ 4.0. $L_{\gamma,min}$ and $L_{\gamma,max}$ correspond to the minimum and maximum luminosity of the sample, respectively. Due to the power-law shape of the LF, changing the upper limits of the integrations has no appreciable effect. The best-fit parameters have been obtained by minimizing the above using the MINUIT minimization package, embedded in ROOT\footnote{\url{https://root.cern/}}. We use the Akaike Information Criterion \citep[AIC;][]{AIC_1974} to evaluate the goodness of fit and we select as the best-fit model the one with the lowest AIC.
 We report both -2ln$L$ and AIC values in Table \ref{tab:PLE}, \ref{tab:PDE} and \ref{tab:LDDE}.

The value of $S$ and its associated 1\,$\sigma$ uncertainty are determined by varying the parameter of interest while allowing all other parameters to float, until a change of $\Delta S = 1$ is reached. This procedure provides an estimate of the 68\,\% confidence interval for the parameter of interest \citep{avni1976}. Although computationally intensive, Eq.~\ref{eq:13} offers the advantage of treating each source's $k$-correction individually. The redshift and the photon index of each source enable us to calculate the corresponding $k$-correction (which is $(1+z)^{\Gamma-2}$ for a power-law spectrum), which is then employed to calculate the corrected rest-frame luminosity from the observed flux.

To visualize whether the best-fit LF provides a good description of the data, we compare the observed redshift, luminosity, photon index and source count distributions against the prediction of the LF. These distributions can be obtained from the LF, respectively, as:
\begin{eqnarray}
\frac{dN}{dz} & = & \int^{\Gamma_{max}}_{\Gamma_{min}} \int^{L_{\gamma,max}}_{L_{\gamma,min}} \lambda(L_{\gamma},\Gamma,z) dL_{\gamma}  d\Gamma \\
\frac{dN}{dL_{\gamma}} & = & \int^{\Gamma_{max}}_{\Gamma_{min}} 
\int^{z_{max}}_{z_{min}} \lambda(L_{\gamma},\Gamma,z) dz d\Gamma \\
\frac{dN}{d\Gamma}  & = &  \int^{L_{\gamma,max}}_{L_{\gamma,min}} 
\int^{z_{max}}_{z_{min}} \lambda(L_{\gamma},\Gamma,z) dz dL_{\gamma}
\end{eqnarray}
\begin{align}
N(>S) =
\int^{\Gamma_{max}}_{\Gamma_{min}} 
\int^{z_{max}}_{z_{min}}  \int^{L_{\gamma,max}}_{L_{\gamma}(z,S)}
 \Phi(L_{\gamma},z)\frac{dN}{d\Gamma}  \frac{dV}{dz} dL_{\gamma} dz d\Gamma
\label{eq:logn}
\end{align}
where $L_{\gamma}(z,S)$ is the luminosity of a source at redshift $z$ having a flux of $S$. Figure \ref{fig:LF_PLE} reports the above distributions for the primarily luminosity evolution model.

Furthermore, to display the LF we rely on the ``N$^{obs}$/N$^{mdl}$'' method \citep{LaFranca1997, Miyaji2001, LaFranca2005, Hasinger2005, Ajello2012}. After obtaining the best-fit LF model, it is possible to determine the value of the observed LF in a given luminosity and redshift bin:
\begin{equation}
\Phi(L_{\gamma,i},z_i) = \Phi^{mdl}(L_{\gamma,i},z_i) \frac {N^{obs}_i}{N^{mdl}_i}
\label{eq:Nobs}
\end{equation}
where $L_{\gamma,i}$ and $z_i$ are the luminosity and redshift of the i$^{th}$
bin and $\Phi^{mdl}(L_{\gamma,i},z_i)$ is the best-fit LF model. $N^{obs}_i$ 
and $N^{mdl}_i$ are the observed and predicted number of FSRQs in that bin, which are used to scale the LF and obtain its observed value. These two approaches — the maximum likelihood (ML) method of \citet{Marshall1983} and the ``N$^{\text{obs}}$/N$^{\text{mdl}}$" estimator — offer a minimally biased estimate of the LF \citep[see also][]{Miyaji2001}.

\subsection{Methods to Construct the LF}
\label{adv_MLMethod}
In this section we discuss different methods that have been used in the literature to generate the LF, which has led to the present-day understanding of the evolution of blazars. The $1/V_{\rm max}$ method in \citet{Schmidt1968} is a binning method to evaluate the LF of a source population. Although widely used to construct LFs, the results depend on the binning interval chosen. Hence, we have not used this method to generate the LF in this study. \cite{Lynden-Bell_C_1971} developed an unbinned non-parametric maximum likelihood method to construct LFs for truncated data, called the $C$-method. \citet{EfronPetrosian_1999} advanced this method by incorporating the C-method for doubly-truncated data, which has also been used to construct the LF of FSRQs \citep[e.g.][]{Singal_FSRQ2014}.
In this study, we have used the parametric unbinned maximum likelihood method introduced by \citet{Marshall1983}. It is a more advanced form of the likelihood approach of \citet{Lynden-Bell_C_1971}, and is best suited for our case. First, since it is a parametric approach, it utilizes an analytical functional form to fit the data. The parameters and function can then be used to calculate various physical quantities of interest. Second, it is an unbinned method and does not depend on binning intervals. Third, it simultaneously fits the local LF and evolution parameters, which enables us to get the whole picture of LF evolution through a mathematical expression. We know that in the case of blazars, the LF evolves with both luminosity and redshift, and our fitting procedure accounts for this. Moreover, our {\it Fermi}-LAT blazar sample is not flux-limited but significance-limited with a strong photon index versus photon flux bias \citep{Abdo2010_fermisurvey}. \citet{Lea2020} performed Monte Carlo simulations to correctly account for these biases in our sample. Together with the parametric form of the LF and accurate description of our sample biases, we are able to generate the FSRQ LF. This makes the method of \citet{Marshall1983}, used in \citet[e.g.,][]{Ajello2012,Ajello2014} the most applicable fitting routine in our study.
\subsection{Uncertainties in the Sky Coverage}
\label{sec:SkyCoverage}
\citet{Lea2020} reported the detection efficiency associated with the sample at $|b|> 20^\circ$, along with its uncertainty. In order to calculate the sky coverage $\Omega(L_\gamma,z,\Gamma)$, we multiply the detection efficiency by the geometric solid angle of the sky surveyed by the telescope for $|b|> 20^\circ$. 

Using the uncertainty associated with the detection efficiency in \citet{Lea2020}, we calculate the upper and lower bounds of the efficiency. We re-run our fitting algorithm using the upper and lower limits of the efficiency, thus obtaining the values of the optimized model parameters for the maximum and minimum. We perform this analysis for the best-fit LF model, thereby obtaining the systematic uncertainty.

\subsection{Variability}
\label{sec:Variability}
Since FSRQs are variable objects (flare $\sim 10\%$ of the time), we consider the effect it may have on our study. First, we have used average quantities throughout our work to mitigate the effects of variability, such as mean flux, mean luminosity and mean photon index. 
\citet{Ajello2012} measure the FSRQ LF using only 1 year of data. If source variability were a problem, we would expect a marked difference between that LF and ours (which relies on 8\,yr of data). Instead, we find (see Section~\ref{sec:results}) that all parameters are consistent within the uncertainty limits, pointing to the fact that the effect of variability is negligible in our study.


\section{Results} \label{sec:results}

\subsection{Primarily Luminosity Evolution}

\begin{figure*}[htb!]
\centering
\includegraphics[width=\textwidth]{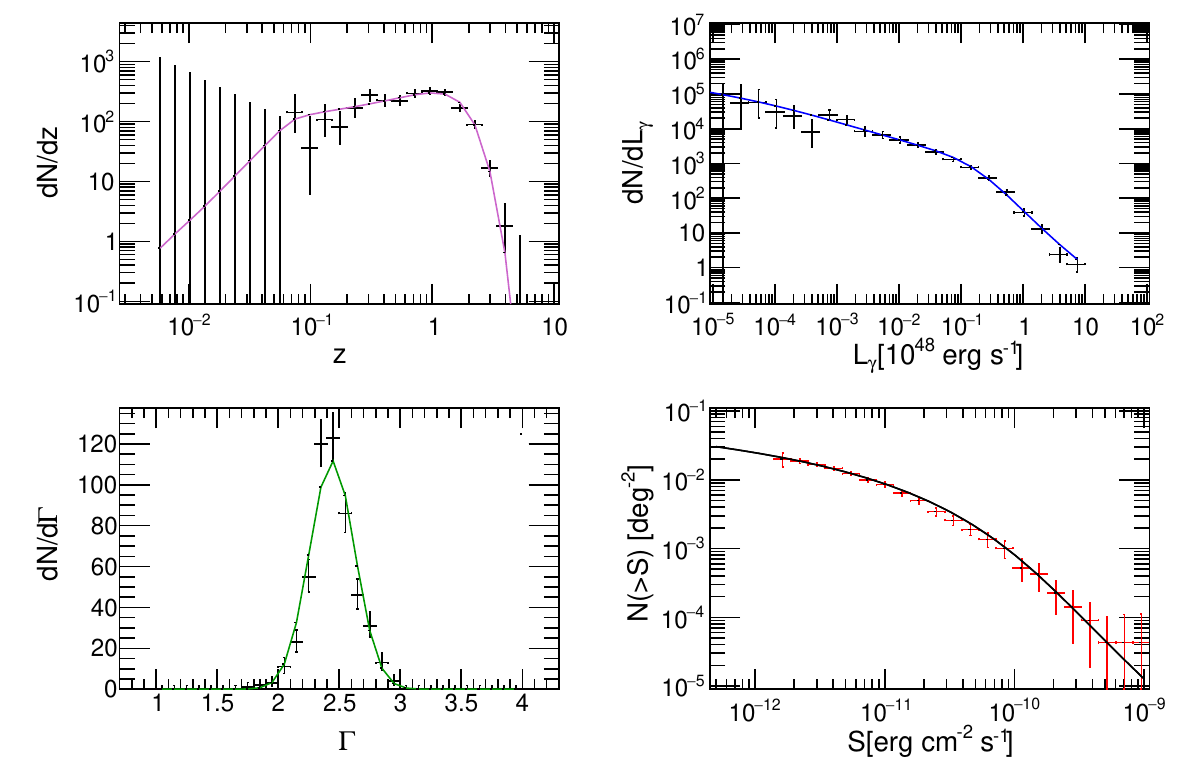}
\caption{Observed redshift (upper left), luminosity (upper right), photon index (lower left),  and intrinsic cumulative source count (lower right) distributions of LAT FSRQs. The solid line is the best-fit line of the PLE model, including the effects of selection bias to compare with the observed data. The error bars represent the statistical uncertainties based on Poisson statistics. In the case of zero sources in a given bin, the $1\sigma$ upper limits are shown \citep[see ][]{Gehrels1986_errorbars}. The red data points (lower right), showing the intrinsic source count distribution, are calculated using the sky coverage.  The corresponding error bars show the propagated errors including the statistical uncertainty and the uncertainty in the sky coverage.}
\label{fig:LF_PLE}
\end{figure*}

\begin{deluxetable*}{llllllllllll}
\tablecaption{Best-fit parameters of the Primarily Luminosity Evolution (PLE) LF.}
\label{tab:PLE}
\tablewidth{0pt}
\tabletypesize{\scriptsize}
\tablehead{
\colhead{$\mu$\textsuperscript{a}} & \colhead{$\sigma$} & \colhead{$\beta$} & \colhead{$A$\textsuperscript{b}} &
\colhead{$L_*$\textsuperscript{c}} & \colhead{$k^*$} & \colhead{$\gamma_1$} & \colhead{$\gamma_2$} &
\colhead{$\xi$} & \colhead{$\tau$} & \colhead{$-2\log L$} & \colhead{AIC}
}
\startdata
2.42 $\pm$ 0.01 & 0.182 $\pm$ 0.005 & 0.026 $\pm$ 0.008 & 5678.0 $\pm$ 249.5 &
0.07 $\pm$ 0.01 & 3.09 $\pm$ 0.55 & 0.69 $\pm$ 0.06 & 3.33 $\pm$ 0.45 &
-0.46 $\pm$ 0.04 & 1.32 $\pm$ 0.09 & -740.242 & -722.242
\enddata
\tablecomments{
\textsuperscript{a} Statistical uncertainties are reported in this table. \\
\textsuperscript{b} In units of $10^{-13}$\,Mpc$^{-3}$\,erg$^{-1}$\,s. \\
\textsuperscript{c} In units of $10^{48}$\,erg\,s$^{-1}$. \\
}
\end{deluxetable*}

In the PLE model, positive evolution is characterized by the value of $k>0$ in Equation \ref{eq:7}, which means that FSRQs were more luminous in the past than they are today. The best-fit parameters are given in Table \ref{tab:PLE}. Figure \ref{fig:LF_PLE} shows the agreement between the observed redshift, luminosity, photon index and source count distributions of \textit{Fermi} FSRQs with our best-fit model, accounting for selection effects.

The value of $k^*=3.09 \pm 0.55$ points towards positive luminosity evolution for the FSRQ sample. The cutoff in the evolutionary parameter is well constrained and is given by $\xi = -0.46 \pm 0.04$. We differentiate Equation \ref{eq:9} to obtain the redshift corresponding to the peak of the distribution. We see that the peak of the luminosity evolution is given by $z_{peak}=-1-k\,\xi$. The peak is calculated to be $z_{peak}= 0.42\pm0.11$, for $L_\gamma = 1.0\times10^{46}$\,erg s$^{-1}$. This is smaller than the value of the redshift peak ($=1.62$) found for the \textit{Fermi} FSRQ sample in \citet{Ajello2012}. However, in that paper, it was observed that the lower luminosity FSRQs ($L_{\gamma}<3.2\times10^{47} \rm erg \, s^{-1}$) in the sample peak at a lower redshift of 1.15. In our sample, the number of FSRQs having lower luminosities than $L_{\gamma}<10^{47} \rm erg \, s^{-1}$ outnumbers the high-luminosity FSRQs and spans two orders of magnitude lower in luminosity than the sample in \citet{Ajello2012}, leading to a lower redshift peak.

\subsection{Primarily Density Evolution}

\begin{figure*}[htb!]
\centering
\includegraphics[width=\textwidth]{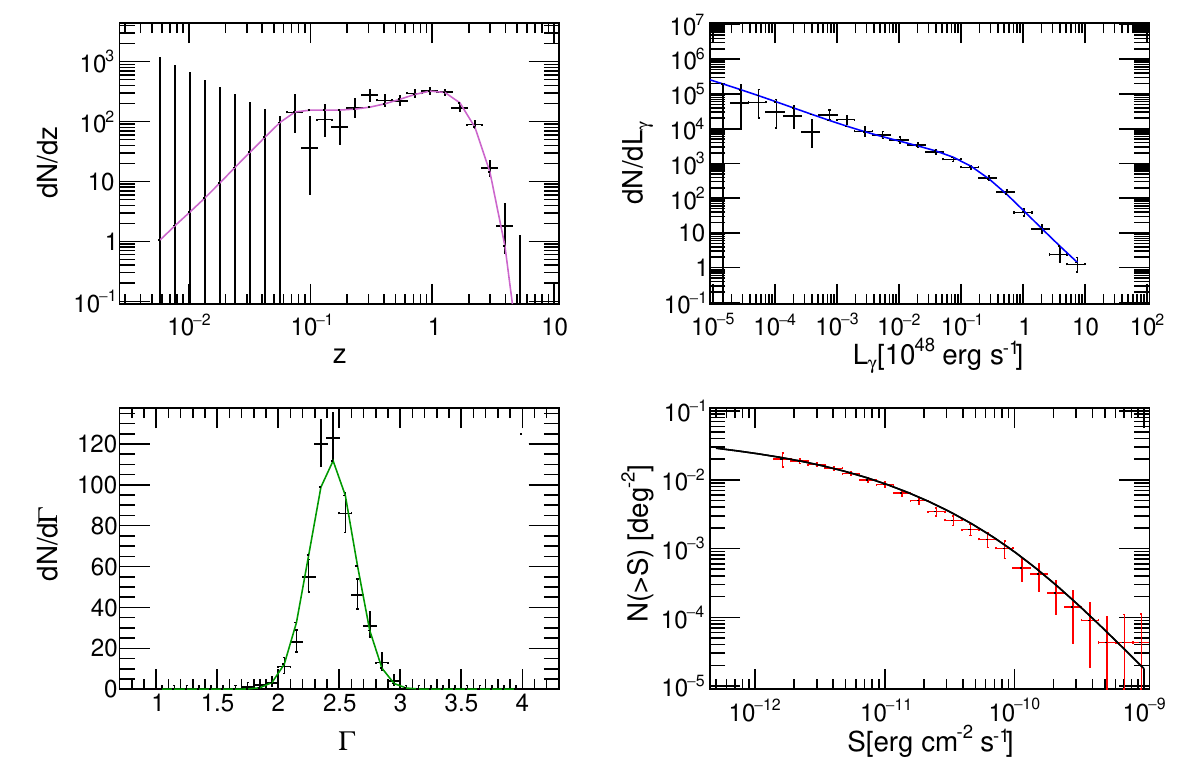}
\caption{Observed redshift (upper left), luminosity (upper right), photon index (lower left),  and intrinsic cumulative source count (lower right) distributions of LAT FSRQs. The solid line is the best-fit line of the PDE model, including the effects of selection bias to compare with the observed data. The error bars represent the statistical uncertainties based on Poisson statistics. In the case of zero sources in a given bin, the $1\sigma$ upper limits are shown \citep[see ][]{Gehrels1986_errorbars}. The red data points (lower right), showing the intrinsic source count distribution, are calculated using the sky coverage.  The corresponding error bars show the propagated errors including the statistical uncertainty and the uncertainty in the sky coverage.}
\label{fig:LF_PDE}
\end{figure*}

\begin{deluxetable*}{llllllllllll}
\tablecaption{Best-fit parameters of the Primarily Density Evolution (PDE) LF.}
\label{tab:PDE}
\tablewidth{0pt}
\tabletypesize{\scriptsize}
\tablehead{
\colhead{$\mu$\textsuperscript{a}} & \colhead{$\sigma$} & \colhead{$\beta$} & \colhead{$A$\textsuperscript{b}} &
\colhead{$L_*$\textsuperscript{c}} & \colhead{$k^*$} & \colhead{$\gamma_1$} & \colhead{$\gamma_2$} &
\colhead{$\xi$} & \colhead{$\tau$} & \colhead{$-2\log L$} & \colhead{AIC}
}
\startdata
2.42 $\pm$ 0.01 & 0.182 $\pm$ 0.005 & 0.026 $\pm$ 0.008 & 832.3 $\pm$ 36.5 &
0.17 $\pm$ 0.06 & 11.26 $\pm$ 1.41 & 0.92 $\pm$ 0.07 & 2.36 $\pm$ 0.17 &
-0.13 $\pm$ 0.01 & 3.62 $\pm$ 0.45 & -747.993 & -729.993
\enddata
\tablecomments{
\textsuperscript{a} Statistical uncertainties are reported in this table. \\
\textsuperscript{b} In units of $10^{-13}$\,Mpc$^{-3}$\,erg$^{-1}$\,s. \\
\textsuperscript{c} In units of $10^{48}$\,erg\,s$^{-1}$. \\
}
\end{deluxetable*}

The positive evolution scenario for the PDE model is characterized by the value of $k>0$ in Equation \ref{eq:6}, which means that FSRQs were more numerous in the past than they are today. The best-fit parameters are shown in Table \ref{tab:PDE}. Figure \ref{fig:LF_PDE} shows how this LF model reproduced the observed data in the redshift, luminosity, photon index and source count distributions of \textit{Fermi} FSRQs with our best-fit parameters, accounting for selection effects.

The value of $k^*=11.26\pm1.41$ implies a very fast positive evolution for the FSRQ sample. The cutoff in the evolutionary parameter is well constrained and is given by $\xi = -0.13 \pm 0.01$. Following a similar method to that of the PLE model, the peak in density evolution is calculated to be $z_{peak}= 0.46\pm0.09$, for $L_\gamma = 1.0\times10^{46}$\,erg s$^{-1}$. This is consistent with the redshift peak found in the PLE case. As shown in Tables \ref{tab:PLE} and \ref{tab:PDE}, the PDE model has a lower value of -2ln$L$ than the PLE ($-747.993$ vs.~$-740.242$, respectively). Moreover, the AIC value is also lower for PDE than for PLE ($-729.993$ vs. $-722.242$, respectively). 

\subsection{Luminosity-Dependent Density Evolution - Best Fit LF}
\label{sec:LDDE}

\begin{figure*}[htb!]
\centering
\includegraphics[width=\textwidth]{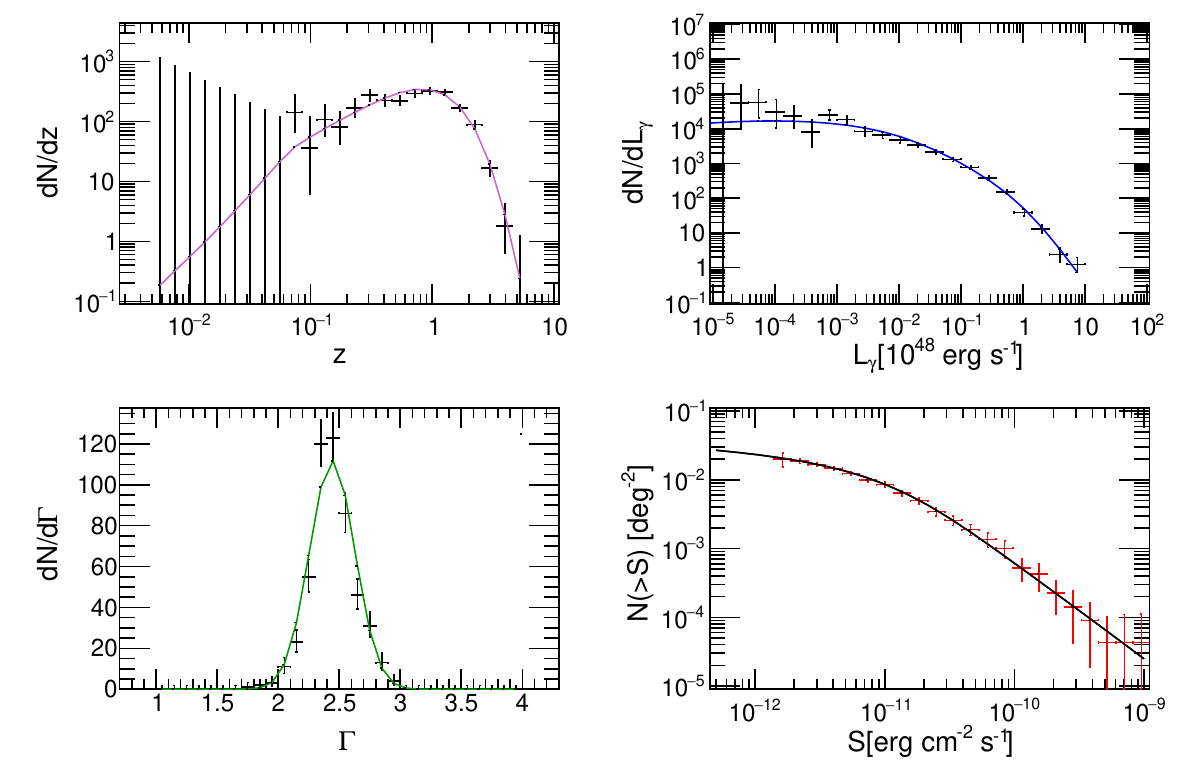}
\caption{Observed redshift (upper left), luminosity (upper right), photon index (lower left),  and intrinsic cumulative source count (lower right) distributions of LAT FSRQs. The solid line is the best-fit line of the LDDE model, including the effects of selection bias to compare with the observed data. The error bars represent the statistical uncertainties based on Poisson statistics. In the case of zero sources in a given bin, the $1\sigma$ upper limits are shown \citep[see ][]{Gehrels1986_errorbars}. The red data points (lower right), showing the intrinsic source count distribution, are calculated using the sky coverage.  The corresponding error bars show the propagated errors including the statistical uncertainty and the uncertainty in the sky coverage.}
\label{fig:LF_LDDE}
\end{figure*}

\begin{deluxetable*}{lllllllllllll}
\tablecaption{Best-fit parameters of the Luminosity-Dependent Density Evolution (LDDE) LF.}
\label{tab:LDDE}
\tablewidth{0pt}
\tabletypesize{\scriptsize}
\tablehead{
\colhead{$\mu$\textsuperscript{a}} & \colhead{$\sigma$} & \colhead{$\beta$} & \colhead{$A$\textsuperscript{b}} &
\colhead{$z^*_c$} & \colhead{$L_*$\textsuperscript{c}} & \colhead{$\gamma_1$} & \colhead{$\gamma_2$} &
\colhead{$\alpha$} & \colhead{$p1$} & \colhead{$p2$} & \colhead{$-2\log L$} & \colhead{AIC}
}
\startdata
2.42 $\pm$ 0.01 & 0.182 $\pm$ 0.005 & 0.025 $\pm$ 0.008 & 18878.4 $\pm$ 830.0 &
2.05 $\pm$ 0.16 & 1.09 $\pm$ 0.52 & 0.29 $\pm$ 0.03 & 1.63 $\pm$ 0.24 &
0.22 $\pm$ 0.01 & 3.5 $\pm$ 1.0 & -9.0 $\pm$ 1.4 & -752.327 & -732.327
\enddata
\tablecomments{
\textsuperscript{a} Total uncertainty is reported as the systematic and statistical uncertainty combined (in quadrature). Systematic uncertainties are derived using the detection efficiency (see Section \ref{sec:SkyCoverage}). \\
\textsuperscript{b} In units of $10^{-13}$\,Mpc$^{-3}$\,erg$^{-1}$\,s. \\
\textsuperscript{c} In units of $10^{48}$\,erg\,s$^{-1}$. \\
}
\end{deluxetable*}

Since \citet{Ajello2012, Ajello2014} showed that a simple PLE or PDE model cannot explain very well the \textit{Fermi} blazar LF, we concentrate on the LDDE model fit. This model is essentially a density evolution model in which objects belonging to different luminosity classes display different redshift peaks, thereby including the effects of luminosity in the evolution. Table \ref{tab:LDDE} gives the best-fit parameters. Figure \ref{fig:LF_LDDE} shows that this model explains the observed redshift, luminosity, photon index, and source count distributions of \textit{Fermi} FSRQs. Comparing Tables \ref{tab:PLE}, \ref{tab:PDE} and \ref{tab:LDDE}, we see that the LDDE model gives the lowest value of -2ln$L$ and AIC. The AIC values are $-722.242$, $-729.993$ and 
$-732.327$ for PLE, PDE and LDDE respectively. The AIC value penalizes each additional parameter added in a model compared to simpler models. We find that the AIC value is least for the LDDE model even after this penalty. This means that the LDDE model is in better agreement with the data compared to the other two models. We define the Akaike weights ($w_i$) as \citep[see][]{Bumham2002_AIC_Model,Wagenmakers2004_AICInference}:
\bbe
w_i = \frac{exp^{-0.5\times \Delta_i}}{\Sigma_{k=1}^{k=N_{model}}exp^{-0.5\times \Delta_k}} \n
\ee
where $\Delta_i$ is:
\bbe
\Delta_i = AIC_i - AIC_{min(best)} \n 
\ee
The evidence ratio is calculated using the Akaike weights, as:
\bbe
{\rm Evidence \, ratio} = \frac{w_j}{w_i}
\ee
which indicates how many times \texttt{Model j} (numerator) is preferred over \texttt{Model i}. Table \ref{Tab:BestModelInference} displays the Akaike weights and the evidence ratios corresponding to the fitted models (PLE and PDE). We find that the LDDE model, according to the data, is $\sim$151 times more likely and $\sim$3 times more likely than the PLE and PDE models, respectively. Moreover, we found that the cross-correlation coefficients between the local LF and evolution parameters for the LDDE model are consistently lower than those of the PDE or PLE models. The cross-correlation coefficients of the LDDE model are at least less than 10\% that of the PLE or PDE models, attesting to the stability of the LDDE model. Thus, we conclude that the LDDE model presented in Equation \ref{eq:10} is the best model to describe the LF of \textit{Fermi} FSRQs.

\begin{table}
\begin{center}
\caption{Comparison of the LF Models.}
\begin{tabular}{ccccc}
\hline
\hline
Model (i)  &  AIC$_i$ & $\Delta_i$ & $w_i$(AIC) & Evidence ratio$^{a}$ \\ 
\hline
\hline
PLE & -722.242   &  10.085  &  0.005  &  151.8  \\
PDE & -729.993   &  2.334  &  0.236  &  3.2  \\
LDDE$^{b}$ & -732.327 (best)  &  0  &  0.759  &  --  \\
\hline
\hline
\multicolumn{5}{l}{
\begin{minipage}{\linewidth} 
\tablenotetext{a}{Evidence ratio (= $w_{\rm LDDE}/w_i$) shows how much more likely the LDDE model is compared to the other models.}
\tablenotetext{b}{The LDDE model has the minimum AIC.}
\end{minipage}
}
\end{tabular}
\label{Tab:BestModelInference}
\end{center}
\end{table}

The best-fit parameters $\gamma_1=0.29 \pm 0.03$ and $\gamma_2=1.63 \pm 0.24$, match well with the parameters given in \citet{Ajello2012} for their smaller sample of \textit{Fermi} FSRQs. The turnover redshift for the most luminous class of our sample ($L_\gamma \sim 10^{48}$) is found to be $z^*_c=$2.05$\pm$0.16. It is important to note that $z_c$ is the redshift at which the evolution changes sign from positive to negative. In the formulation of LDDE, we have scaled this parameter as dependent on the luminosity through a power-law relation (see Equation \ref{eq:alpha}). The power-law index of the evolution of the redshift is $\alpha$. We obtain the value $\alpha = 0.22 \pm 0.01$, which is similar to that obtained in \citet{Ajello2012} ($\alpha_{\rm Ajello\_2012} = 0.21 \pm 0.03$). Our uncertainty ranges of both the systematic and statistical uncertainty are considerably less than those of \citet{Ajello2012}; refer to Appendix $A.1$ of that paper. Since we have many more objects in the $z$ range of $2-3$, and leading up to $z\sim 4$, this provides further confidence in our parameter estimation.

\begin{figure}[htb!]
\centering
\includegraphics[width=\columnwidth]{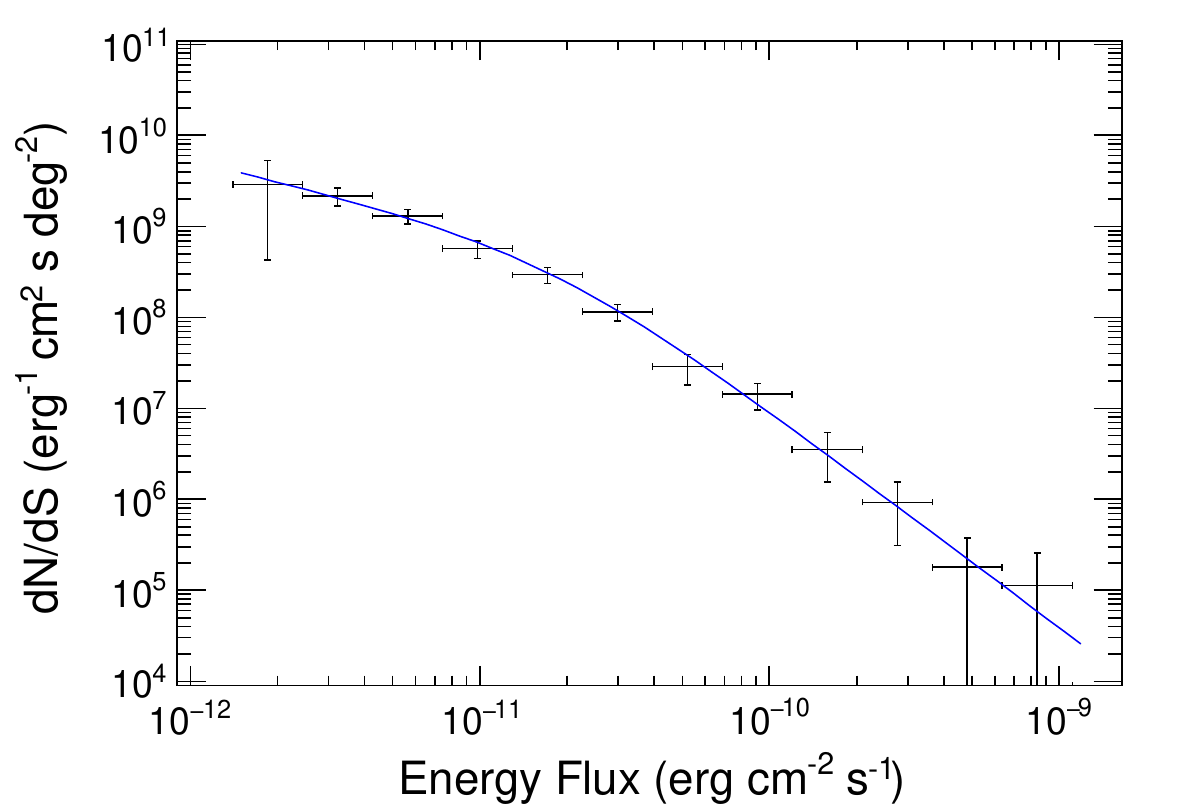}
\caption{The intrinsic differential source count distribution (\textit{logN-logS}) of FSRQs plotted against the energy flux. The solid blue line represents the intrinsic \textit{logN-logS} as derived using the LDDE model. The data points are calculated using Equation \ref{eqn:dNdS}. The error bars show the propagated errors including the statistical uncertainty and the uncertainty in the sky coverage.}
\label{fig:dNdS}
\end{figure}

Figure \ref{fig:dNdS} shows the intrinsic differential source count distribution (\textit{logN-logS}) of FSRQs. The intrinsic number of sources in a particular flux interval is calculated using the equation:
\bbe
\frac{dN}{dS} = \frac{1}{\Omega(S_i)} \frac{N_i}{\Delta S_i}
\label{eqn:dNdS}
\ee
where $N_i=$ the number of FSRQs in the $i$-th flux bin, $S_i =$ the $i$-th energy flux, $\Delta S_i = $ the flux interval centered at $S_i$, and $\Omega(S_i) =$ the sky coverage at flux $S_i$. The solid blue line indicates the source count distribution predicted by the LDDE model.

\begin{figure*}[htb!]
\centering
\includegraphics[width=0.8\textwidth]{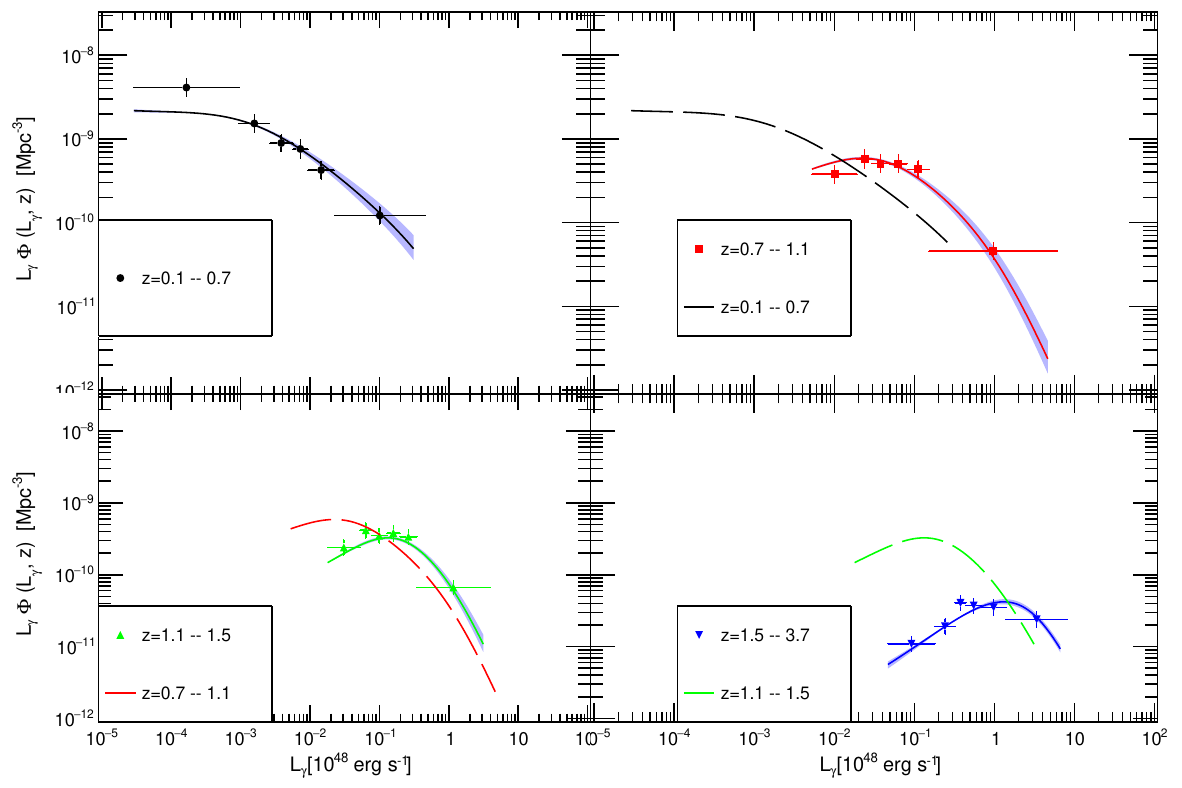}
\caption{The LF of \textit{Fermi} FSRQs subdivided into four bins in redshift, constructed using the $N_{obs}/N_{mdl}$ method. The solid lines depict the best-fit model (LDDE) in each redshift bin, along with the space density in the previous bin given in dashed line for comparison. The systematic uncertainty shown in the blue shaded region was derived using the detection efficiency (see Section \ref{sec:SkyCoverage}).}
\label{fig:LF_Evol_L}
\end{figure*}

\begin{figure*}[htb!]
\centering
\includegraphics[width=0.8\textwidth]{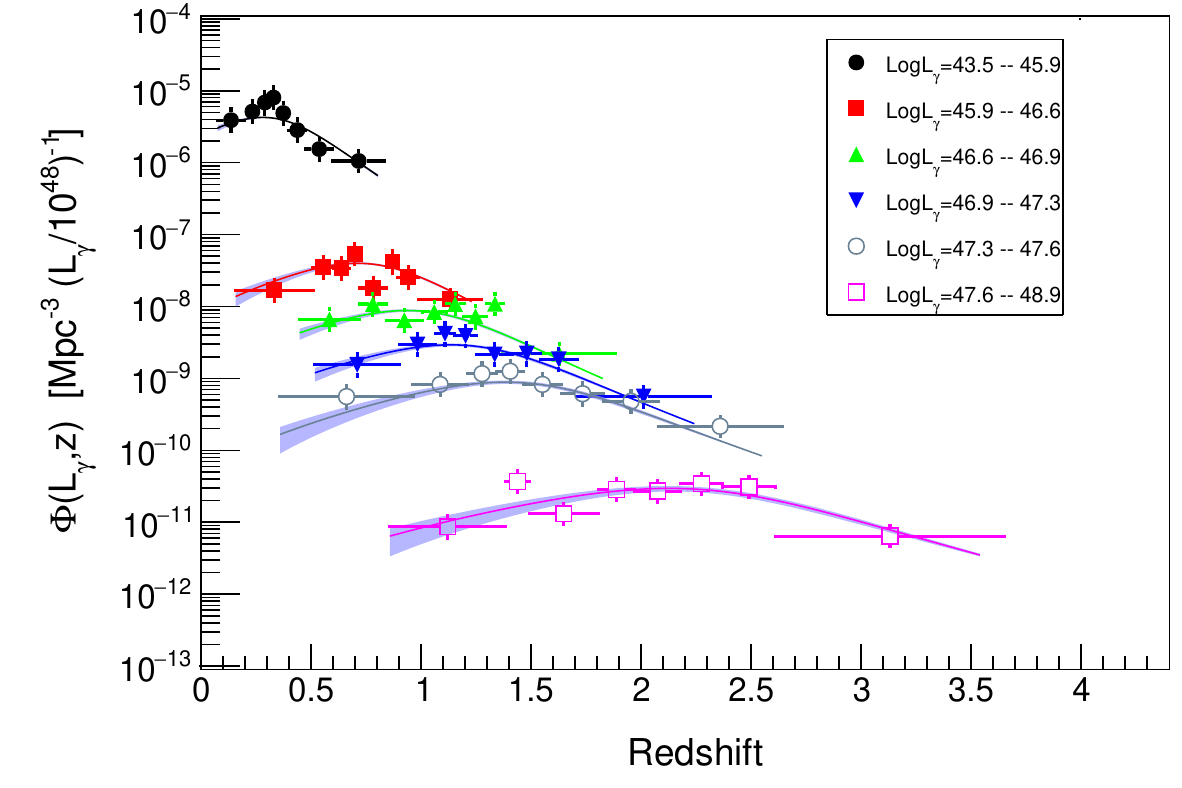}
\caption{The evolution of \textit{Fermi} FSRQs subdivided into six bins in luminosity, constructed using the $N_{obs}/N_{mdl}$ method by employing the best-fit (LDDE) model. The peak in evolution is at higher redshift for a more luminous class; that is, this plot shows the luminosity dependence of evolution in density. The systematic uncertainty shown in the blue shaded region was derived using the detection efficiency (see Section \ref{sec:SkyCoverage}).}
\label{fig:LF_Evol_z}
\end{figure*}

In Figure \ref{fig:LF_Evol_L} we divide our sample into four redshift bins, with roughly equal numbers of sources in each bin, to illustrate the evolution of the LF in each bin. Figure \ref{fig:LF_Evol_z} shows a similar division of the sample into six luminosity bins to visualize the luminosity-dependent evolution of LF. Both of these representations have been constructed using Equation \ref{eq:Nobs}. The evolving LF can be seen as shifting of the peak and the shape of the plots in different redshift bins in Figure \ref{fig:LF_Evol_L}. The major evolution episode seems to have occurred for $z<1.1$ and $z>1.5$. The main findings of this exercise are:
\begin{itemize}
    \item The number density of sources having $\log L=46$ seems to remain the same for $z<1.1$, and decreases for higher $z$.
    \item Between $1.1<z<1.5$ the number density of sources having log $L=$ 47 appears to remain the same, whereas those with log $L=$ 48 continue to increase.
    \item For $z>1.5$, the number density of sources having log $L=$ 47 decreases sharply, whereas those with log $L=$ 48 seem to remain the same.
\end{itemize}
The evolution of the redshift peak is parameterized by $\alpha=0.22\pm0.01$, and is visually depicted in Figure \ref{fig:LF_Evol_z}, where the peak of evolution of different luminosity classes occurs at different epochs. More specifically, the peak of LF occurs at higher redshift for more luminous objects. 

\subsection{The Local LF}
\label{sec:localLF}

\begin{figure*}[htb!]
\centering
\includegraphics[width=0.8\textwidth]{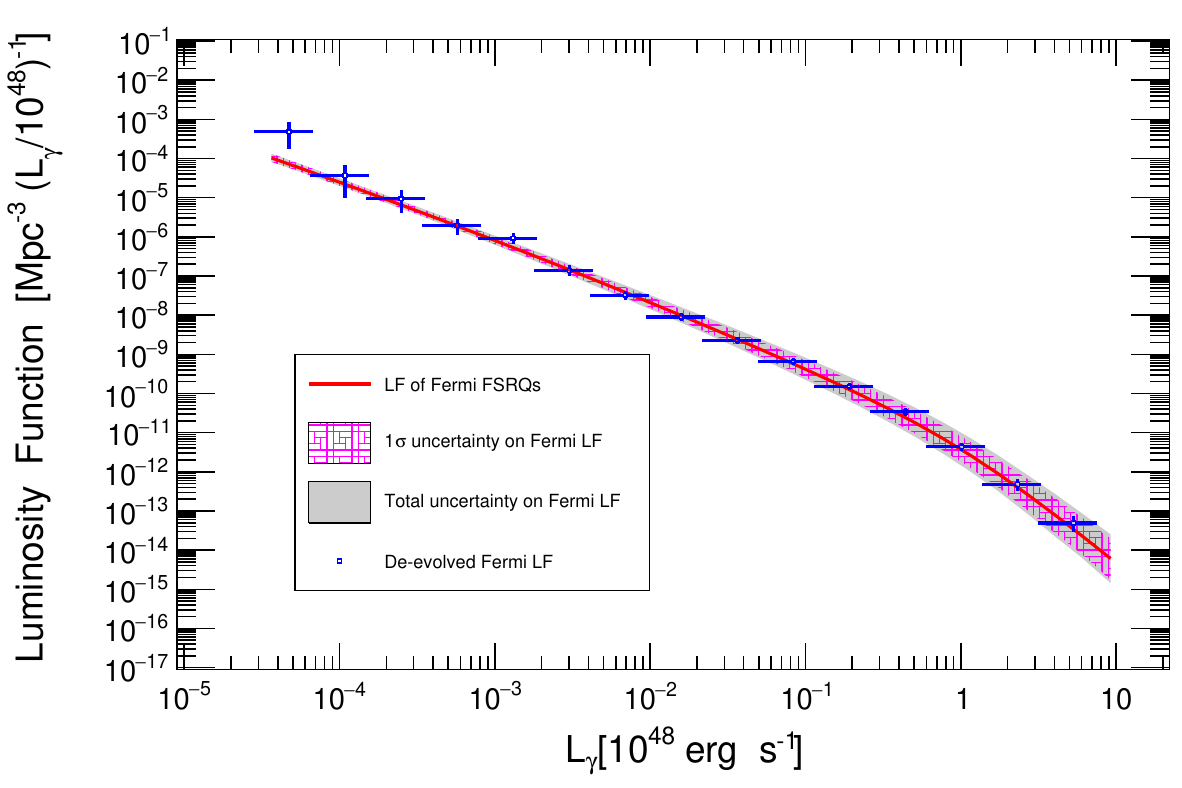}
\caption{The local LF of \textit{Fermi} FSRQs, de-evolved at $z=0$. The blue points are the calculated values from the $1/V_{\rm max}$ method. The pink hatched area is the $\pm1\sigma$ uncertainty of the LF. The total uncertainty is shaded in grey. The total uncertainty is calculated by combining the statistical ($1\sigma$ uncertainty) and systematic uncertainties (see \ref{sec:SkyCoverage}). The red line depicts the best-fit line of the local LF.}
\label{fig:LF_DeEvol}
\end{figure*}

The local LF refers to the luminosity function at $z=0$. It is obtained by de-evolving the densities (or luminosities) according to the best-fit model, to a redshift of $0$. Since our best-fit model is the LDDE model, we need to take into account the effect of luminosity dependence while de-evolving the densities. We employ the $1/V_{\rm max}$ method \citep{Schmidt1968} including the correction for the luminosity-dependent evolution, proposed by \citet{DellaCeca2008} as:
\bbe
V_{\rm max} = \int^{z_{max}}_{z_{min}} \Omega(L_i,z) \frac{e'(z,L_i)}{e'(z_{min},L_i)}
  \frac{dV}{dz}dz
\ee
where $L_i$ is the source luminosity, $\Omega(L_i,z)$ is the sky coverage,
$z_{max}$ is the redshift above which the source drops out of the survey,
and $e'(z,L_i)$ is the evolution term of Eq.~\ref{eq:10} normalized
(through $e'(z_{min},L_i)$) at the redshift $z_{min}$ to which
the LF is to be de-evolved. 

To also capture the model uncertainty in reconstructing the LF, we conduct a Monte Carlo simulation. We bootstrap parameters from the covariance matrix of the best-fit LDDE model described in Section \ref{sec:LDDE}. The resampled parameters are then used to calculate the $\pm1\sigma$ uncertainty of the LF at redshift zero, as shown in Figure \ref{fig:LF_DeEvol} in the pink hatched shaded area. We estimate the total uncertainty (shaded in grey) by combining (in quadrature) the $\pm1\sigma$ uncertaintiy and the systematic uncertainty (see \ref{sec:SkyCoverage}). We see in Figure \ref{fig:LF_DeEvol} that the local LF can be described by a double power law, steepening at higher luminosities. This change occurs near $L_\gamma> 10^{47} \rm erg \, s^{-1}$.

\section{Discussion} 
\label{sec:conc}

\begin{figure}[htb!]
\centering
\includegraphics[width=\columnwidth]{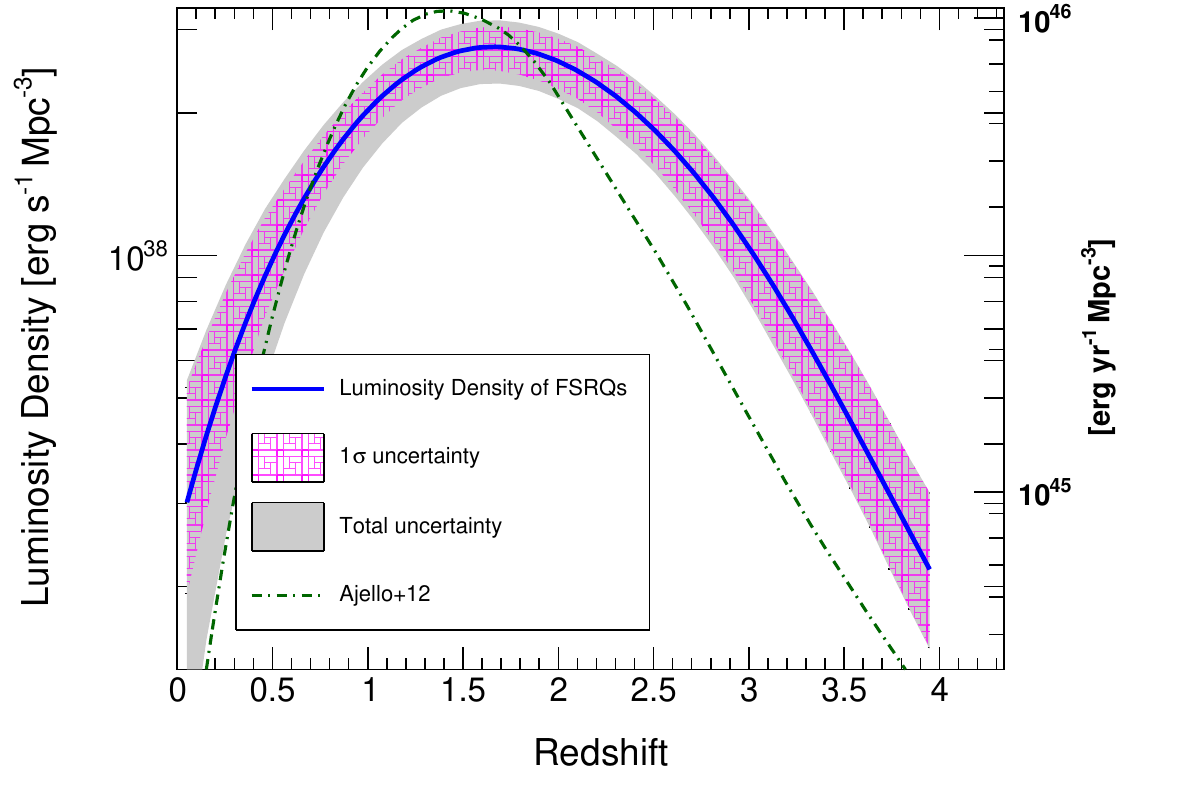}
\caption{The luminosity density of \textit{Fermi} FSRQs. The best-fit line of our model is represented by the blue line, surrounded by the 1$\sigma$ uncertainty region shaded in pink hatched pattern and the total uncertainty shaded in grey. The green dash-dotted line shows the luminosity density of FSRQs obtained using best-fit parameters from \citet{Ajello2012}.}
\label{fig:LumDens}
\end{figure}

\begin{figure}[htb!]
\centering
\includegraphics[width=\columnwidth]{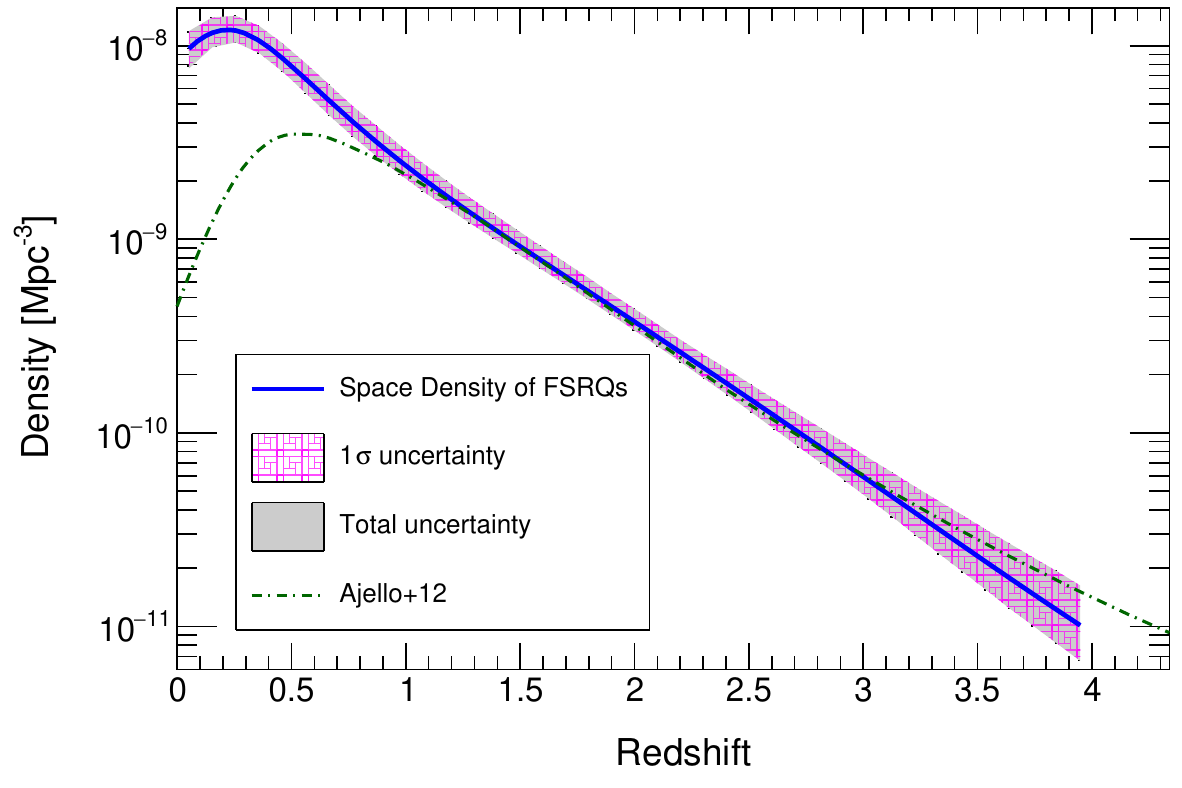}
\caption{The space density of \textit{Fermi} FSRQs. The best-fit line of our model is represented by the blue line, surrounded by the 1$\sigma$ uncertainty region shaded in pink hatched pattern and the total uncertainty shaded in grey. The green dash-dotted line shows the space density of FSRQs obtained using best-fit parameters from \citet{Ajello2012}.}
\label{fig:NumDens}
\end{figure}

In this work, we have utilized the largest FSRQ dataset --- by a factor of 4 --- to date to construct the $\gamma$-ray LF of FSRQs. Our sample consists of 519 FSRQs detected by {\it Fermi}-LAT at high Galactic latitude ($|b|>20^\circ$) and between  100\,MeV and 1\,TeV. By fitting analytical models that best represent our statistical data, we have constrained the LF of FSRQs. In this section, we compare our best-fit LF to previous studies and discuss the major implications of our best-fit LF.

Our investigation confirms that the evolution of FSRQs is best described by a LDDE model. This agrees with previous studies of X-ray selected radio-quiet AGNs \citep{Ueda2003,Hasinger2005,LaFranca2005,Wall2008}, $\gamma$-ray selected EGRET blazars \citep{NarumotoTotani2006} and \textit{Fermi} blazars \citep{Ajello2012, Ajello2014}. 
When comparing the best-fit LDDE parameters  with those of \citet{Ajello2012} we see that the parameters are compatible within uncertainties. 

The local LF, i.e. the LF at $z=0$, is best represented by a double power law. This shape is also seen in radio-quiet and radio-loud AGNs \citep{DunlopPeacock1990,Ueda2003,Hasinger2005,NarumotoTotani2006}. As we can see in Figure \ref{fig:LF_DeEvol}, there is a flattening of the slope at the faint-end of the luminosity. This could be caused by beaming which is known to flatten the LF at low luminosity \citep[see e.g.][]{UrryShafer1984,Urry1991,Ajello2012}.

Figure \ref{fig:LumDens} displays the luminosity density of FSRQs, as computed using the best-fit LF. The figure shows that the luminosity density reached a maximum between $z=1-2$. This epoch is also the peak of the star formation rate of the universe \citep[e.g.,][]{HopkinsBeacom2006,MadauDickinson2014,DaCunha2023}, thereby suggesting a connection between the galactic nucleus and the host. For comparison, we also include the luminosity density of FSRQs computed in \citet{Ajello2012} in Figure \ref{fig:LumDens}, and find agreement with our work.

The LDDE model suggests that the space density of objects of different luminosities peaks at different redshifts, which can be seen in Figure \ref{fig:LF_Evol_z}. This means that objects of higher luminosity reached their peak space density much earlier in cosmic time than low luminosity objects. This paucity of higher luminosity objects in recent times is true for all classes of AGNs \citep[e.g.,][]{Cowie1999,Hasinger2005} and was also shown for \textit{Fermi} FSRQs in \citet{Ajello2012}. The formation and fueling of SMBHs have a major role to play in the evolution of blazars, with more massive black holes powering luminous FSRQs at large redshifts \citep{Volonteri2011, Sbarrato2015_SMBH}. The spin of black holes is also connected to the accretion disk and jet of blazars \citep{Blandford1977, ZhangSpin2024}. This means that the reduction in the number of powerful FSRQs in the recent universe may be due to the downsize of fast-spinning black holes \citep{Dubois2014_BHSpin,Wang2019}. The evolution of accretion activity of black holes can also lead to this paucity of FSRQs in the nearby universe \citep[e.g.][]{Merloni2004}. Quasar activity could be ignited by major galaxy mergers, feeding the central black holes with more gas \citep{Sanders1988, DiMatteo2005}. The reduction of galaxy mergers and scarcity of gas in the later universe can throttle quasar activity leading to a decrease in the number of FSRQs \citep[e.g.][]{Hopkins2008,KulkarniLoeb2012}.

Figure \ref{fig:NumDens} shows the number density of FSRQs calculated from our best-fit LF. We see the overall number density peak is in a redshift range of $0.2-0.4$, followed by a rapid decline governed by the parameter $p_2=-9.0\pm1.4$. \citet{Ajello2012} found that the space density of FSRQs peak at $z\sim 0.5$. This can be attributed to the fact that the minimum luminosity of our sample ($L_{\gamma,min}=2.9\times10^{43}$\,erg s$^{-1}$) is 2 orders of magnitude less than the sample used in \citet{Ajello2012} ($L_{\gamma,min}=3.9\times10^{45}$\,erg s$^{-1}$). The inclusion of more objects of lower luminosity drives the peak of space density to lower redshifts ($z<1$).


\begin{figure}[htb!]
\centering
\includegraphics[width=\columnwidth]{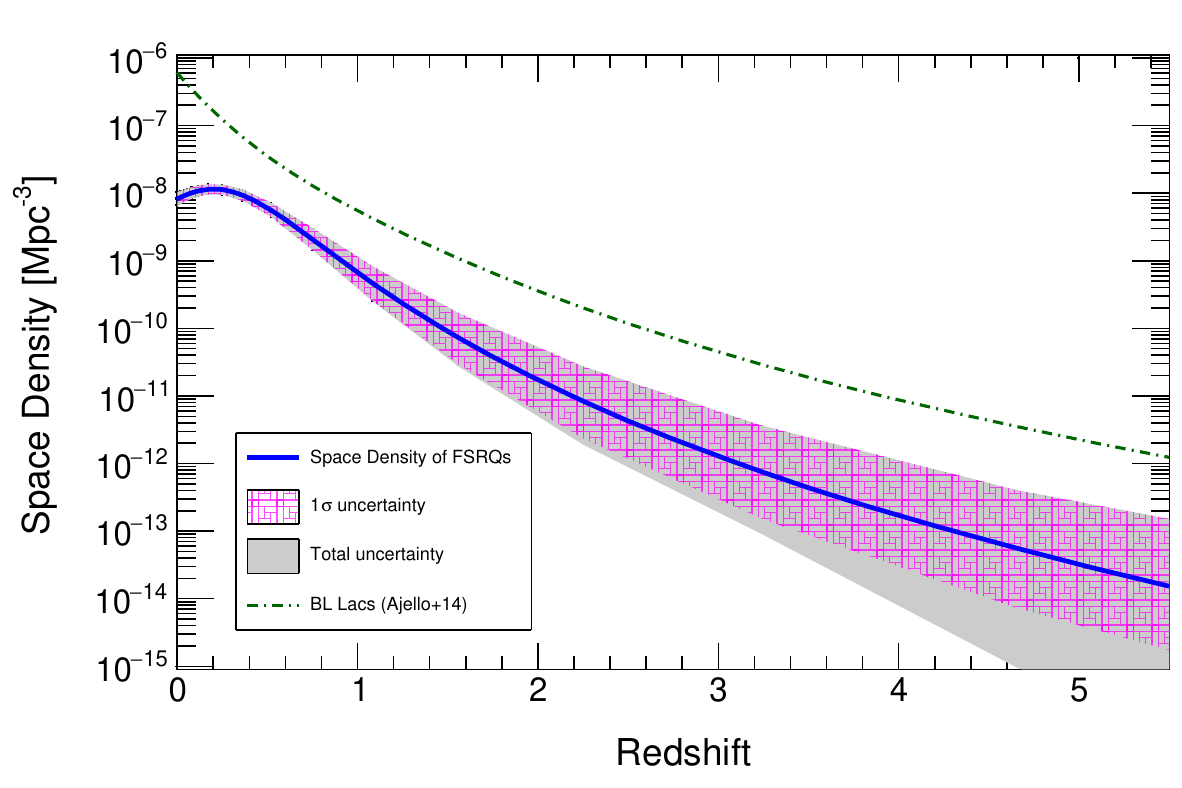}
\caption{The space density of \textit{Fermi} FSRQs in the low luminosity range: $\log L_\gamma$ (erg s$^{-1}$) = 43.5 $-$ 45.6. The best-fit line of our model is represented by the blue line, surrounded by the 1$\sigma$ uncertainty region shaded in pink hatched pattern and the total uncertainty shaded in grey. The green dash-dotted line shows the space density of BL Lacs obtained using best-fit model and parameters from \citet{Ajello2014}, within the same low luminosity range.}
\label{fig:BLL_low}
\end{figure}

\begin{figure}[htb!]
\centering
\includegraphics[width=\columnwidth]{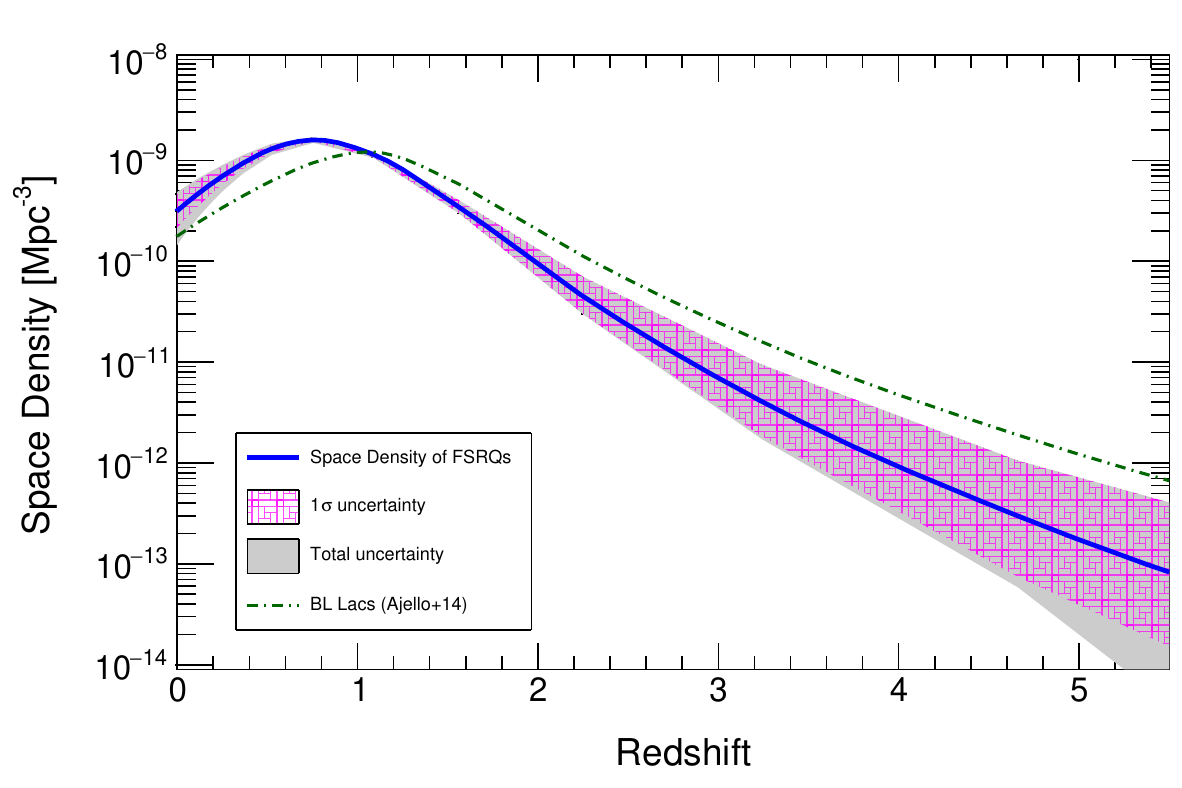}
\caption{The space density of \textit{Fermi} FSRQs in the intermediate luminosity range: $\log L_\gamma$ (erg s$^{-1}$) = 45.8 $-$ 47.2. The best-fit line of our model is represented by the blue line, surrounded by the 1$\sigma$ uncertainty region shaded in pink hatched pattern and the total uncertainty shaded in grey. The green dash-dotted line shows the space density of BL Lacs obtained using best-fit model and parameters from \citet{Ajello2014}, within the same intermediate luminosity range.}
\label{fig:BLL_mid}
\end{figure}

\begin{figure}[htb!]
\centering
\includegraphics[width=\columnwidth]{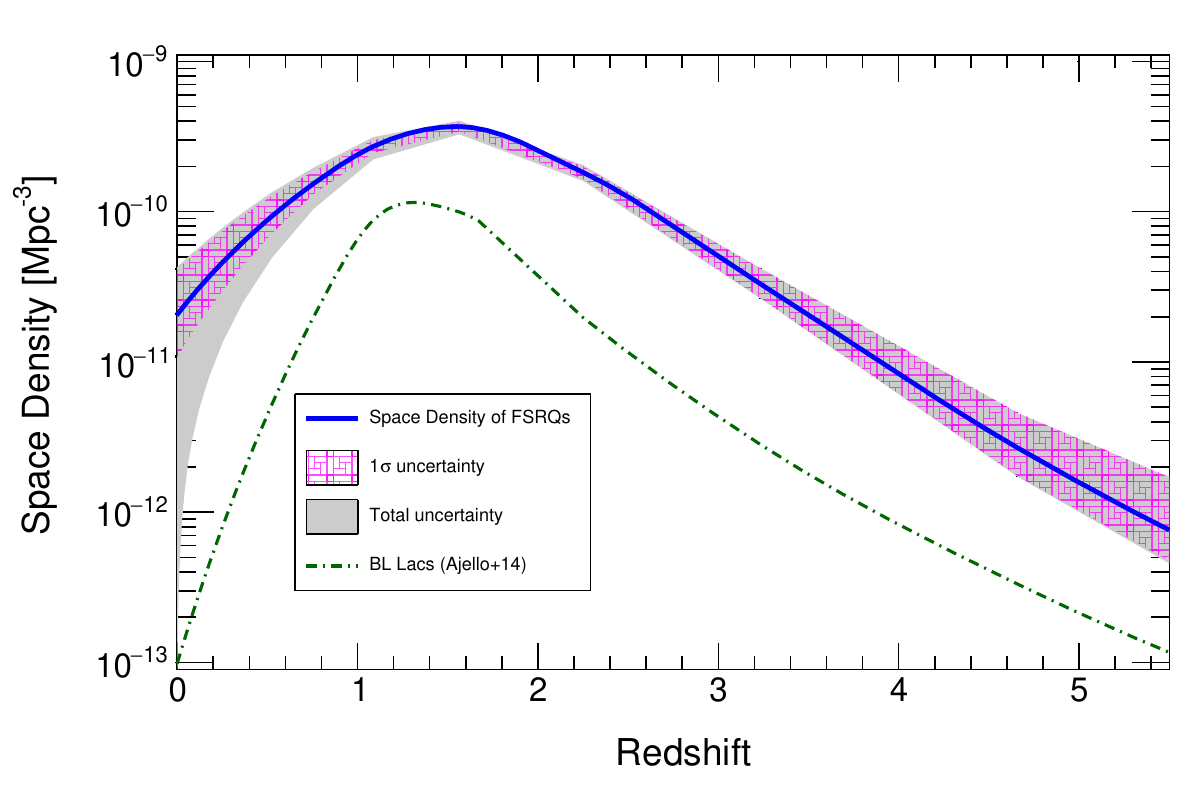}
\caption{The space density of \textit{Fermi} FSRQs in the high luminosity range: $\log L_\gamma$ (erg s$^{-1}$) = 47.2 $-$ 48.4. The best-fit line of our model is represented by the blue line, surrounded by the 1$\sigma$ uncertainty region shaded in pink hatched pattern and the total uncertainty shaded in grey. The green dash-dotted line shows the space density of BL Lacs obtained using best-fit model and parameters from \citet{Ajello2014}, within the same high luminosity range.}
\label{fig:BLL_high}
\end{figure}

We compare the space density of our sample to that one of \citet{Ajello2012} and find agreement for $z\gtrsim1$. Below that redshift, our measured space density is one order of magnitude larger than the one measured by \citet{Ajello2012}. This is certainly due to the fact that our sample reaches much lower luminosities than that of \citet{Ajello2012}.

\subsection{Comparison with the Evolution of BL Lacs}
\label{sec:Con_BLLCompare}

Figure \ref{fig:BLL_low} shows the space density of low-luminosity FSRQs ($\log L_\gamma$(\rm erg s$^{-1}$) = 43.5 $-$ 45.6) compared to that of BL Lacs of similar luminosity as derived by \citet{Ajello2014}. The two classes follow similar evolutions up to low redshift where the BL Lac evolution remains negative, while the FSRQ one switches to positive.

Figure \ref{fig:BLL_mid} shows a comparison of the space densities of FSRQs and BL Lacs of intermediate luminosity ($\log L_\gamma$(\rm erg s$^{-1}$) = 45.8 $-$ 47.2). The space density of FSRQs seem to peak at $z\sim0.8$, whereas that of BL Lacs peak at $z\sim 1.2$. Quite surprisingly, we find that the evolution of these intermediate luminosity blazars is nearly identical.

Figure \ref{fig:BLL_high} shows a comparison of the space densities of FSRQs and BL Lacs of high luminosity ($\log L_\gamma$(\rm erg s$^{-1}$) = 47.2 $-$ 48.4). We find again a similarity in the shape of the evolution of both sub-classes.


The surprising similarity in the space density of $\gamma$-ray selected FSRQs and BL Lacs may have two underlying reasons. First, the BL Lac sample may be contaminated by FSRQs  incorrectly classified as BL Lacs. This can happen if FSRQs have a jet nearly perfectly aligned with our line of sight making the detection of broad emission lines in their optical spectra impossible.
Since the classification is carried out mainly on the basis of equivalent width, an increase in the flux of the jet emission can swamp out prominent emission lines \citep[e.g.,][]{Ruan2014, Mishra2021}. Moreover, at high redshifts ($z>0.7$), the most prominent emission lines (such as $\rm H\alpha$) can be redshifted to infrared wavelengths (outside the optical window) and not detected, leading to the misclassification of FSRQs as BL Lacs \citep[e.g.,][]{DElia2015}. FSRQs mistaken as BL Lacs are commonly referred to as `masquerading BL Lacs' \citep{padovani2019} and a few of them have been identified so far \citep{Ghisellini2012_blueFSRQ,Padovani2012,Giommi2013,Rajagopal2020,Rajguru2024}. It is easy to misclassify such luminous `blue FSRQs' as BL Lacs, thereby contaminating the BL Lac sample and giving it an implicit bias towards a FSRQ-like evolution.

On the other hand, FSRQs and BL Lacs may  share the same evolution because in reality they represent the same class of objects artificially divided on the basis of observational appearances (presence or lack of broad emission lines). The different accretion rate \citep[e.g.,][]{FermiBlazarDivide2009} and emission mechanism (whether or not there is an external inverse Compton component) can be explained if there is a strong genetic link between FSRQs and BL Lacs \citep[as proposed by e.g.][]{Vagnetti1991_LinkBLL, cavaliere2002,boettcher2002}, with BL Lacs 
representing the gas-starved radiatively inefficient phase of blazars, while FSRQs have abundant cold gas fueling a radiatively efficient accretion disk, which in turn sustains strong broad emission lines and a prominent external radiation field. The difference in the physical properties of FSRQs and BL Lacs can also be due to difference in magnetic field strength \citep[e.g.][]{Mondal2019}, black hole spin \citep[e.g.][]{Bhattacharya2016,Gardner2018} and beaming effect \citep[e.g.][]{FanZhang2003}, where BL Lacs possess stronger magnetic fields, lower spins and radiatively inefficient accretion disks compared to FSRQs. \citet{Padovani1992_Relation} suggested that FSRQs and BL Lacs could be explained as different beaming scenarios in high- and low-luminosity radio galaxies. Thus both the classes of blazar may display the same trend in the evolution of number densities despite having different observational characteristics. In any case this is probably the first time such a similarity in the LF of the two classes is noted and it is partly due to the larger number of low luminosity FSRQs in this sample.

\section{Summary}
\label{sec:summary}

This work presents the $\gamma$-ray LF of FSRQs by using a sample of 519 sources detected by \textit{Fermi}-LAT, in the energy range of 100 MeV - 1 TeV. The main findings of this study are summarized below:

\begin{enumerate}
    \item We use the largest sample of FSRQs (519 sources) available in the literature to construct their LF. The sample has luminosities down to $\log L_\gamma=43.46$, probing a large number of low luminosity objects. The large redshift span of the sample enables us to study the evolution of FSRQs up to $z\sim4$. All 519 FSRQs in our sample have redshifts measured in the literature. In addition, a robust determination of detection efficiency was also carried out for this sample by \citet{Lea2020}, providing us with the quantitative description of the survey and its selection biases.
    \item We find that FSRQs follow a density evolution that depends on the luminosity, i.e., the LDDE model. There exists a characteristic redshift ($z_c(L_\gamma)$) at which the LF transitions from a positive to negative evolution. This redshift, when the LF peaks, depends on the luminosity of the FSRQ, indicating that more luminous FSRQs reached their maximum LF at earlier cosmic times (see Figure \ref{fig:LF_Evol_z}). Figure \ref{fig:LF_Evol_L} shows the evolution of the comoving number density in different redshift bins.  This shows that FSRQs having higher luminosities were more common in the past, until the peak of the LF, after which they decreased in number.
    \item The local LF is determined by de-evolving the LF according to our best-fit model. We see that the local LF follows a double power-law shape, which is typical of beamed sources \citep{UrryShafer1984}.
    \item The luminosity density of FSRQs peaks at a redshift range of $z\sim 1-2$. The number density of FSRQs is maximum in the redshift range of $z\sim0.2-0.4$, followed by a rapid decline in number at higher redshifts.
    \item One of the most intriguing findings of this work is the surprising similarity in the number density of FSRQs and BL Lacs. The evolution of both the classes of blazars follow the same pattern in evolution with an offset in the peak of LF. This can be due to the BL Lac sample having prominent FSRQ contamination, or both blazar classes having the same intrinsic evolutionary trend with difference in their physical properties only.

\end{enumerate}

\vspace{5cm}
We thank the referee and the statistics reviewer for comments that helped in improving the manuscript. We acknowledge the NASA grant 80NSSC24K0284 for this project. This research has used redshifts available in the online data base ZBLLAC (\url{https://web.oapd.inaf.it/zbllac/index.html}).

%






\appendix
\restartappendixnumbering
\section{The Sample of Flat-Spectrum Radio Quasars}
\label{sec:appx_sample}

Table \ref{tab:FSRQ_List} provides the composition of the $\gamma$-ray sample of FSRQs used in our study.

\begin{deluxetable*}{llllllllllll}[h!]
\tablecaption{Composition of the Gamma-Ray Sample of Flat-Spectrum Radio Quasars$^{a}$.}
\label{tab:FSRQ_List}
\tablewidth{0pt}
\tabletypesize{\scriptsize}
\tablehead{
\colhead{Source Name} & \colhead{4FGL Name} & \colhead{RA$^{b}$ (J2000)} & \colhead{DEC$^{b}$ (J2000)} & \colhead{GLAT$^{b}$} & \colhead{eflux$^{c}$} & \colhead{eflux uncertainty$^{c}$} & \colhead{CLASS} &
\colhead{Redshift} & \colhead{PL Index} & \colhead{PL Index uncertainty} & \colhead{Counterpart Name}
}
\startdata
PS J0001.4$+$2113 & 4FGL J0001.5$+$2113 & 0.38 & 21.21 & $-$40.1676 & 1.82$\times 10^{-11}$ & 7.58$\times 10^{-13}$ & FSRQ &  1.106 &  2.65 &  0.02 &  TXS 2358$+$209 \\
PS J0004.3$-$4737 & 4FGL J0004.4$-$4737 & 1.10 & $-$47.62 & $-$67.5378 & 6.66$\times 10^{-12}$ & 4.79$\times 10^{-13}$ & FSRQ & 0.88 & 2.43 & 0.05 & PKS 0002$-$478 \\
PS J0005.9$+$3825 & 4FGL J0005.9$+$3824 & 1.49 & 38.40 & $-$23.6133 & 8.62$\times 10^{-12}$ &6.08$\times 10^{-13}$ & FSRQ & 0.229 & 2.62 & 0.05 & S4 0003$+$38 \\ 
PS J0010.3$+$2045 & 4FGL J0010.6$+$2043 & 2.65 & 20.73 & $-$41.1282 & 2.88$\times 10^{-12}$ & 4.96$\times 10^{-13}$ & FSRQ & 0.6 & 2.39 & 0.15 & TXS 0007$+$205 \\
PS J0010.6$-$3025 & 4FGL J0010.6$-$3025 & 2.66 & $-$30.42 & $-$80.4823 & 4.71$\times 10^{-12}$ & 4.54$\times 10^{-13}$ & FSRQ & 1.19 & 2.42 & 0.05 & PKS 0008$-$307 \\
\enddata
\tablecomments{
\textsuperscript{a} Table \ref{tab:FSRQ_List} is published in its entirety in the machine-readable format. A portion is shown here for guidance regarding its form and content. \\
\textsuperscript{b} In units of deg. \\
\textsuperscript{c} Energy flux over 100 MeV - 1 TeV (in units of $\rm erg\, cm^{-2}\, s^{-1}$). \\
}
\end{deluxetable*}





\end{document}